# Generation of spin-polarized currents via cross-relaxation with dynamically pumped paramagnetic impurities

*Carlos A. Meriles[1], Marcus W. Doherty[2]*

[1]Dept. of Physics, CUNY-City College of New York, New York, NY 10031, USA.

[2]Laser Physics Centre, Research School of Physics and Engineering, Australian National University, Canberra, Australian Capital Territory 0200, Australia

**Abstract**

Key to future spintronics and spin-based information processing technologies is the generation, manipulation, and detection of spin polarization in a solid state platform. Here, we theoretically explore an alternative route to spin injection via the use of dynamically polarized nitrogen-vacancy (NV) centers in diamond. We focus on the geometry where carriers and NV centers are confined to proximate, parallel layers and use a 'trap-and-release' model to calculate the spin cross-relaxation probabilities between the charge carriers and neighboring NV centers. We identify near-unity regimes of carrier polarization depending on the NV spin state, applied magnetic field, and carrier *g*-factor. In particular, we find that unlike holes, electron spins are distinctively robust against spin-lattice relaxation by other, unpolarized paramagnetic centers. Further, the polarization process is only weakly dependent on the carrier hopping dynamics, which makes this approach potentially applicable over a broad range of temperatures.

Electrical spin injection into a non-magnetic material has become today an established experimental technique[1,2]. The typical geometry is that of a layered 'spin valve', where the non-magnetic system is surrounded by ferromagnetic electrodes. Carriers are spin polarized in one of the electrodes and transported through the non-magnetic medium via the applied electric field; the spin polarization persisting after the transit is determined magnetoresistively, by using the second electrode as a spin analyzer. This form of spin injection and detection has been demonstrated in multiple material platforms, including inorganic semiconductor heterostructures[3-5] and organic heterojunctions[6,7]. Other spin injection and transport schemes that do not rely on the use of ferromagnetic electrodes have been investigated as well. Examples are optical spin injection[8], and injection of spin polarization from point contacts and quantum dots[9,10]. More recently, spin filtering of an electrical current in a semiconductor has been attained by exploiting spin-selective carrier recombination near engineered defects[11].

Here, we theoretically investigate an alternate route to spin injection in which carriers spin-polarize by Overhauser cross-relaxation with dynamically pumped paramagnetic centers. We focus on the case of a diamond-based structure designed to bring charge carriers and optically pumped NV centers into close proximity with each other. We assume a 'trap-and-release' model of charge transport — in which carriers hop between localized trap sites — and find regimes of efficient carrier polarization at moderate magnetic fields and over a broad range of hopping rates. Distinct spin dynamics are predicted for carriers exhibiting *g*-factors other than that of the NV, including the appearance of carrier spin hyperpolarization at high magnetic field. Further, our results suggest that electrons might be more resilient than holes to depolarization via background, unpolarized spin-1/2 centers. Finally, we discuss some possible physical realizations in the form of layered diamond heterotructures or conducting films over NV-rich surfaces.

We consider charge carriers with spin number $I = 1/2$ and g-factor $g_I$ confined to a conductive plane located a distance $d \sim 1 - 3$ nm from a thin layer of paramagnetic centers with spin number $S$ and g-factor $g_S$ (Fig. 1). Carrier diffusion across the conductive plane is governed by a trap-and-release dynamics characterized by a thermally activated hopping time $\tau_c$, and the carrier and trap concentrations are assumed to be such that correlation effects can be ignored[12]. Neglecting for now spin-lattice contributions, the combined system of carrier and center spins is described by the simplified Hamiltonian

$$H = H_I + H_S + H_{dip}(t) \qquad (1)$$

where $H_I$ and $H_S$ are respectively the carrier and center spin Zeeman and crystal field Hamiltonians, and $H_{dip}(t)$ is the time-dependent carrier-center dipolar interaction. For a given carrier-center pair, $H_{dip}(t)$ takes the form

$$H_{dip}(t) = F_0(t)\left(I_z S_z - \frac{1}{4}(I_+ S_- + I_- S_+)\right) + F_1(t)(I_+ S_z + I_z S_+) +$$

$$+F_1^*(t)(I_- S_z + I_z S_-) + F_2(t)I_+ S_+ + F_2^*(t)I_- S_- \qquad (2)$$

where we use the standard nomenclature to denote spin operators. In Eq. (2), the functions $F_j(t), j = 0,1,2$ of the dipolar interspin vector are time-dependent due to carrier hopping between traps. They are given by $F_0(t) = \frac{k}{r^3}(1 - 3\cos^2\theta)$, $F_1(t) = -\frac{3}{2}\frac{k}{r^3}\sin\theta \cos\theta\, e^{i\varphi}$, and $F_2(t) = -\frac{3}{4}\frac{k}{r^3}\sin^2\theta\, e^{2i\varphi}$, where $(r, \theta, \varphi)$ denote the spherical coordinates of the interspin vector in a reference frame whose z-axis coincides with the NV symmetry axis, $k \equiv \mu_0 g_I g_S \mu_B^2/(4\pi\hbar^2)$, $\mu_B$ the Bohr magneton, $\mu_0$ the magnetic permeability of vacuum, and $\hbar$ Planck's constant divided by $2\pi$.

We consider carriers interacting with a layer of engineered NV centers ($S=1$). This defect exhibits a triplet ground state with a room-temperature zero-field splitting $\Delta = 2.87$ GHz. Upon optical excitation, initialization into the $|m_S = 0\rangle$ state takes place on a ~1 μs time scale[13]. As the carrier hops from one trap to the next, it experiences a time-dependent coupling to the polarized NV, which causes spin relaxation. We assume sufficiently dilute concentrations of carriers and centers such that interspin interactions are weak and clustering can be ignored. In this limit, first-order perturbation theory yields the cross-relaxation transition probabilities $W_{\alpha\beta} = \frac{1}{t\hbar^2}\left|\int_0^t \langle\alpha|H_{dip}(t')|\beta\rangle e^{-i\omega_{\alpha\beta}t'} dt'\right|^2$ between states $|\alpha\rangle = \left|m_S^{(\alpha)}, m_I^{(\alpha)}\right\rangle$ and $|\beta\rangle = \left|m_S^{(\beta)}, m_I^{(\beta)}\right\rangle$ of the interacting $I$-$S$ system[14]. Denoting time averages with brakets and spatial averages with overbars, such that $\langle F_j(t) F_j^*(t+\tau)\rangle = \overline{|F_j|^2} e^{-|\tau|/\tau_c}$ for $j=0\ldots 2$, we find the unit time probabilities

$$W_0 = \frac{\xi_0^{(NV)} \tau_c}{1 + \left(\left(\omega_{NV}^{(-1)} - \omega_I\right)\tau_c\right)^2} \; ; \qquad \widetilde{W}_0 = \frac{\xi_0^{(NV)} \tau_c}{1 + \left(\left(\omega_{NV}^{(+1)} - \omega_I\right)\tau_c\right)^2} \; ;$$

$$W_1 = \frac{\xi_1^{(NV)} \tau_c}{1 + (\omega_I \tau_c)^2} \; ; \qquad (3)$$

$$W_2 = \frac{\xi_2^{(NV)} \tau_c}{1 + \left(\left(\omega_{NV}^{(-1)} + \omega_I\right)\tau_c\right)^2} \; ; \qquad \widetilde{W}_2 = \frac{\xi_2^{(NV)} \tau_c}{1 + \left(\left(\omega_{NV}^{(+1)} + \omega_I\right)\tau_c\right)^2} \; ;$$

where $\xi_0^{(NV)} = \hbar^2 \overline{|F_0|^2}/4, \xi_1^{(NV)} = 2\hbar^2 \overline{|F_1|^2}, \xi_2^{(NV)} = 4\hbar^2 \overline{|F_2|^2}$, $\omega_{NV}^{(\pm 1)} = \omega_{crys} \pm |g_S|\mu_B B/\hbar$, $\omega_{crys} = 2\pi\Delta$, $\omega_I = g_I \mu_B B/\hbar$, and $\tau_c$ is the correlation time characterizing the carrier hopping dynamics. The subscripts $j=0,1,2$ are a short-hand notation for $j = \left|m_S^{(\alpha)} - m_S^{(\beta)} + m_I^{(\alpha)} - m_I^{(\beta)}\right|$, and it is assumed that the magnetic field direction coincides with the NV axis.

To model the dynamics of spin polarization we make use of detailed population balance and derive the system master equation for the carrier spin polarization $\langle I \rangle$

$$\frac{d\langle I \rangle}{dt} = -\left(\langle I \rangle - I^{(e)}\right)\left(2W_1(1-P_0) + (W_2+W_0)(1-P_{+1}) + (\widetilde{W}_2+\widetilde{W}_0)(1-P_{-1}) + \frac{1}{T_{1I}}\right)$$

$$-\left(\langle S_{0,-1}\rangle - S^{(e)}_{0,-1}\right)(W_2 - W_0) - \left(\langle S_{0,+1}\rangle - S^{(e)}_{0,+1}\right)(\widetilde{W}_0 - \widetilde{W}_2), \quad (4)$$

where $P_{m_S}$ denotes the (normalized) NV population in state $|m_S\rangle$, and $\langle S_{m_S^{(\alpha)}, m_S^{(\beta)}} \rangle \equiv \frac{1}{2}\left(P_{m_S^{(\alpha)}} - P_{m_S^{(\beta)}}\right)$ corresponds to the virtual spin-1/2 magnetization between NV states $\left|m_S^{(\alpha)}\right\rangle$ and $\left|m_S^{(\beta)}\right\rangle$; we use the superscript $(e)$ to indicate thermal equilibrium and assume that the carrier spin lifetime due to processes other than cross-relaxation with the NVs is $T_{1I}$. Using Eq. (4) one can derive the carrier polarization in the stationary state (i.e., when $\frac{d\langle I\rangle}{dt} = 0$) under different pumping conditions. For example, when NVs are continously illuminated with strong green light, we have $P_0 \approx 1$ and the corresponding steady-state carrier magnetization takes the form

$$\langle I \rangle_0 = -\frac{1}{2}\frac{\left((W_2-W_0)-(\widetilde{W}_2-\widetilde{W}_0)\right)}{\left(W_2+W_0+\widetilde{W}_2+\widetilde{W}_0+1/T_{1I}\right)}, \quad (5)$$

where we neglect the thermal magnetization of NVs and carriers.

To explicitly determine the carrier spin polarization we examine the case of a [100] diamond crystal — for which the NV axis angle with the normal to the conductive surface is $\psi = 54.7°$, Fig. 1 — and derive the set of transition probability amplitudes $\left(\xi_0^{(NV)}, \xi_1^{(NV)}, \xi_2^{(NV)}\right) = (\pi \hbar^2 k^2 \sigma_{NV}/d^4)(0.094, 0.31, 0.59)$ where $\sigma_{NV} = 10^{12}$ cm$^{-2}$ is the NV surface density[15,16]. The polarization $P_I^{(0)} = 2\langle I \rangle_0$ as determined from Eq. (5) is presented in Fig. 2a as a function of the magnetic field amplitude $B$ and hopping time $\tau_c$ for the case of a carrier with g-factor $g_I = -2$ and intrinsic lifetime $T_{1I} = 10$ μs. We find almost complete (negative) spin polarization in the vicinity of 50 mT, where $|\omega_I| \sim \omega_{NV}^{(-1)}$ and $P_I^{(0)} \approx -1/(1 + 1/W_2 T_{1I})$. The sharp dependence on $B$ is a consequence of the relatively short value assumed for $T_{1I}$. Further, since $W_2 T_{1I} \propto T_{1I} \tau_c / \left(1 + \left(\left(\omega_{NV}^{(-1)} + \omega_I\right)\tau_c\right)^2\right)$, the range of fields where dynamic carrier polarization is efficient narrows as the hopping time $\tau_c$ becomes longer, in agreement with Fig. 2a. In the regime of short correlation times ($\tau_c < 0.1$ ns for the present conditions), $W_2 T_{1I} \to 0$ and thus $P_I^{(0)} \to 0$.

If the crystal contained NVs aligned with the other three possible NV directions, then carrier cross-relaxation with these misaligned centers must also be considered, though the anticipated carrier polarization enhancement is small. Additionally, there have been recent reports of fabricated crystals with near perfect preferential alignment of NVs[17,18], in which case these results are directly applicable.

While optical excitation drives the NV into $|m_S = 0\rangle$, effective pumping into $|m_S = -1\rangle$ could be attained by concatenating light excitation and microwave pulses resonant with $\omega_{NV}^{(-1)}$. Using Eq. (4), the steady-state magnetization reads in this case

$$\langle I \rangle_{-1} = \frac{1}{2} \frac{(W_2 - W_0)}{(W_2 + 2W_1 + W_0 + 1/T_{1I})}, \tag{6}$$

plotted in Fig. 2b for the same range of magnetic fields and hopping times. Interestingly, we observe a sign reversal of the polarization relative to that in Fig. 2a, which could be exploited to gate the carrier spin polarity[9]. By the same token, pumping into $|m_S = +1\rangle$ yields

$$\langle I \rangle_{+1} = -\frac{1}{2} \frac{(\widetilde{W}_2 - \widetilde{W}_0)}{(\widetilde{W}_2 + 2W_1 + \widetilde{W}_0 + 1/T_{1I})}, \tag{7}$$

presented in Fig. 2c. Unlike the former two cases, the polarization is low within the observed range of magnetic fields and correlation times due to the mismatch between $\omega_{NV}^{(+1)}$ and $\omega_I$.

Besides NV centers, diamond hosts a variety of paramagnetic defects that could also be exploited for carrier polarization. An example is the P1-center, a spin-1/2 defect formed by substitutional nitrogen. When P1-centers polarized into $m_S = +1/2$ serve as the source of carrier spin polarization, population balance leads to the steady-state formula[19]

$$\langle I \rangle_{+1/2} = -\frac{1}{2} \frac{(V_2 - V_0)}{\left(V_2 + 2V_1 + V_0 + 1/T_{1I}^{(a)}\right)}, \qquad (8)$$

where, as before, we neglect thermal polarization. The P1-induced transition probabilities $V_j$ can be obtained from the expressions for $W_j$ in Eq. (3) with the correspondence $\omega_{NV}^{(-1)} \to \omega_{P1} = g_{P1}\mu_B B/\hbar$ and $\left(\xi_0^{(NV)}, \xi_1^{(NV)}, \xi_2^{(NV)}\right) \to \left(\xi_0^{(P1)}, \xi_1^{(P1)}, \xi_2^{(P1)}\right) = \left(\xi_0^{(NV)}/2, \xi_1^{(NV)}/4, \xi_2^{(NV)}/2\right)$. The plot of $\langle I \rangle_{+1/2}$ as a function of the magnetic field and correlation time is presented in Fig. 2d assuming conditions identical to those used for the NVs. We find near uniform carrier polarization once $B$ and $\tau_c$ exceed a minimum threshold (~10 mT and ~1 ns, respectively), which reflects the condition $g_I = g_{P1}$ assumed in this case. This type of response could be attained, e.g., by dynamically polarizing the ensemble of P1 centers via neighboring NV centers[20].

Since P1 concentrations in diamond are typically comparable or larger than that of the NVs, a question of interest is whether carrier spin relaxation induced by *unpolarized* P1 centers can offset the polarization gain due to dynamically pumped NVs. After extending Eq. (3) to include contributions from both polarized NVs and unpolarized P1-centers, we obtain the steady-state formula

$$\langle I \rangle_0 = -\frac{1}{2} \lambda_0 \frac{\left((W_2 - W_0) - (\widetilde{W}_2 - \widetilde{W}_0)\right)}{\left(W_2 + W_0 + \widetilde{W}_2 + \widetilde{W}_0\right)}, \qquad (9)$$

where $\lambda_0 = 1/\left(1 + \left(V_2 + 2V_1 + V_0 + 1/T_{1I}^{(a)}\right)/\left(W_2 + W_0 + \widetilde{W}_2 + \widetilde{W}_0\right)\right)$ is a 'leakage factor', and we assume NV initialization into $m_S = 0$. For carriers with $g$-factor $g_I = -2$ (electrons), a near-depolarization-free regime is found over a reasonably broad range of correlation times and magnetic fields (Fig. 3a). This result reflects the condition $\lambda_0 \sim 1/(1 + V_0/W_2)$ near $B \sim 50$ mT, where $W_2$ and $V_0$ are the dominant relaxation channels via NV and P1 centers, respectively. Since $\xi_2^{(NV)} \sim 10 \xi_0^{(P1)}$, we have for equal concentrations of NV and P1 centers $W_2 \sim 10 V_0$ near 50 mT and thus $\lambda_0 \sim 1$ in the vicinity of this field, in agreement with the numerical calculation. A different response is found if $g_I = 2$ (holes) because carrier spin relaxation via NVs and P1s is then dominated by $W_0$ and $V_2$, respectively. In this case $W_0 \sim 2V_2/5$ near 50 mT, thus leading to a peak leakage factor $\lambda_0 \sim 1/(1 + V_2/W_0) \sim 0.3$, further reduced in Fig. 3b by intrinsic carrier spin relaxation.

The dynamic polarization of carriers whose g-factors differ from 2 or -2 is worth exploring on its own. Neglecting for now relaxation through P1 centers, Fig. 4b shows the calculated steady-state polarization as a function of the carrier g-factor and magnetic field for NVs pumped into $m_S = 0$ and assuming $\tau_c = 1$ ns. Besides the interval around 50 mT (Figs. 2a and 2b), the case $|g_I| < 2$ leads to a complementary regime of dynamic polarization at high magnetic field originating from the condition $\omega_{NV}^{(-1)} = -\omega_I$ (Fig. 4a). The polarization rate of carriers with positive (negative) g-factor is dominated here by $W_2$ ($W_0$) thus leading to negative (positive) $P_I$, opposed to that found near 50 mT. As $|g_I|$ approaches $|g_{NV}|$ the range of magnetic fields where carriers polarize increases: An example is shown in Fig. 4b for carriers with Lande factor $g_I = -1.85$, where spin alignment is attained near 1.35 T over an interval exceeding 200 mT (ten times larger than at low field). If unpolarized P1 centers are present, arguments similar to those above lead to $\lambda_0|_{g_I=-1.85} \sim 1/(1 + V_0/W_0) \sim 0.6$ when $\sigma_{P1} = \sigma_{NV}$. Therefore, substantial carrier spin polarization should be attainable even in the presence of background paramagnetic centers. A similar analysis in the case $|g_I| > 2$ leads to high-field polarization when $\omega_{NV}^{(+1)} \sim \omega_I$ (Fig. 4b). Note that for $g \geq 3$, a steep polarization sign change takes place with a small variation in the magnetic field, which could facilitate carrier polarization switching[10].

The geometry in Fig. 1 could be attained, for example, within a diamond heterostructure by successively growing δ-doped layers of different composition so as to stack polarization and conduction planes separated by intrinsic diamond layers; *p*-type[21] or *n*-type[22] conduction is possible with a suitable selection of the dopant (e.g., boron[15] or phosphorous[23], respectively). Detection of the carrier spin polarization could be carried out via the inverse effect, namely, by monitoring the fluorescence of a separate group of probe, unpolarized NVs exposed to polarized carriers. The technology required to produce this type of heterostructure is available and has been utilized recently to fabricate diamond FET transistors[21,24,25]. In these devices conduction between source and drain takes place around a 1 nm thick δ-doped layer separated from the gate electrode by a 1 nm layer of intrinsic diamond. Interestingly, a similar gate geometry has been exploited to tune the *g*-factor of carriers confined to a semiconductor quantum well[26]. This approach to *g*-factor control could perhaps be extended to diamond, a low-spin-orbit-coupling material where, nonetheless, virtually no spin transport experiments have yet been reported. Of note, the relative diamond crystal orientation plays only a secondary role. The latter is demonstrated in Fig. 4d where we plot the steady-state polarization as a function of $\psi$, the angle between the NV axis and the normal to the conductive plane (Fig. 1). We observe a smooth transformation governed by the dependence of the probability amplitudes $\xi_0^{(NV)}$ and $\xi_2^{(NV)}$ on $\psi$ (Fig. 4c). This change, however, is small and thus insufficient to significantly alter the spin carrier response as calculated above.

Besides the diamond-based heterostructures introduced above, near-surface NV centers could be used, e.g., to polarize carriers propagating through thin films of organic conductors or other low-spin-orbit-coupling materials (e.g., graphene or silicon) deposited on the diamond surface. Likewise, the use of other paramagnetic centers is conceivable: Worth highlighting among the latter are the di-vacancy and silicon-vacancy centers in SiC[27,28], substitutional phosphorous[29] and bismuth[30] in silicon, and rare-earth ions in wide-bandgap semiconductors[31], all of which have proven to polarize upon light excitation at room temperature. Since attaining the longest defect spin lifetimes is not necessarily a requisite for spin injection, some of these systems may prove advantageous as compared to diamond, either because surface defects can be created more efficiently or closer to the surface (e.g., Ce impurities in YAG), or because the light power required to initialize them is less than for NVs. By the same token, and while *optical illumination* is the most common form of spin initialization, other systems may be better suited for *electrical* spin pumping. One illustration is the case of self-interstitial Ga atoms in GaNAs, which self-polarize at room temperature upon circulation of a current[32].

C.A.M. acknowledges support from the National Science Foundation through grant NSF-1314205. M.W.D. acknowledges support from the Australian Research Council through grant DP120102232. We are thankful to J. Wrachtrup, F. Jelezko, P. Neumann, J.A. Reimer, N.B. Manson, and L.T. Hall for helpful discussions.

**Meriles and Doherty, Fig. 1**

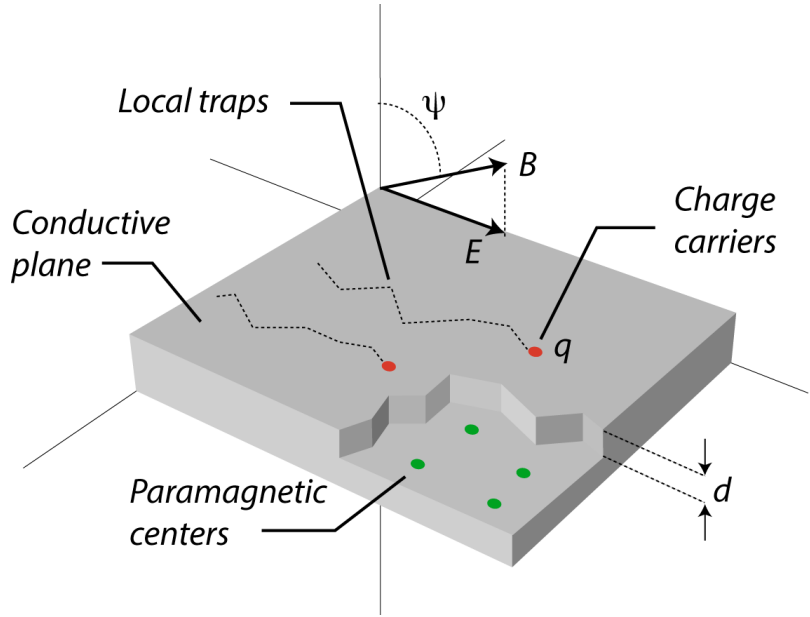

**Fig. 1**: System geometry. Charge carriers with *g*-factor $g_I$ (red dots) are confined to a plane located a distance $d$ from a layer of externally pumped paramagnetic centers (green dots). A 'trap-and-release' model is assumed in which carriers hop from one trap to the next on the conducting plane at a rate $1/\tau_c$, where $\tau_c$ denotes the carrier correlation time. The applied magnetic field $B$ forms an angle $\psi$ with the carrier plane normal; an in-plane electric field $E$ (optional) is applied to drag carriers across the NV-rich zone.

**Meriles and Doherty, Fig. 2**

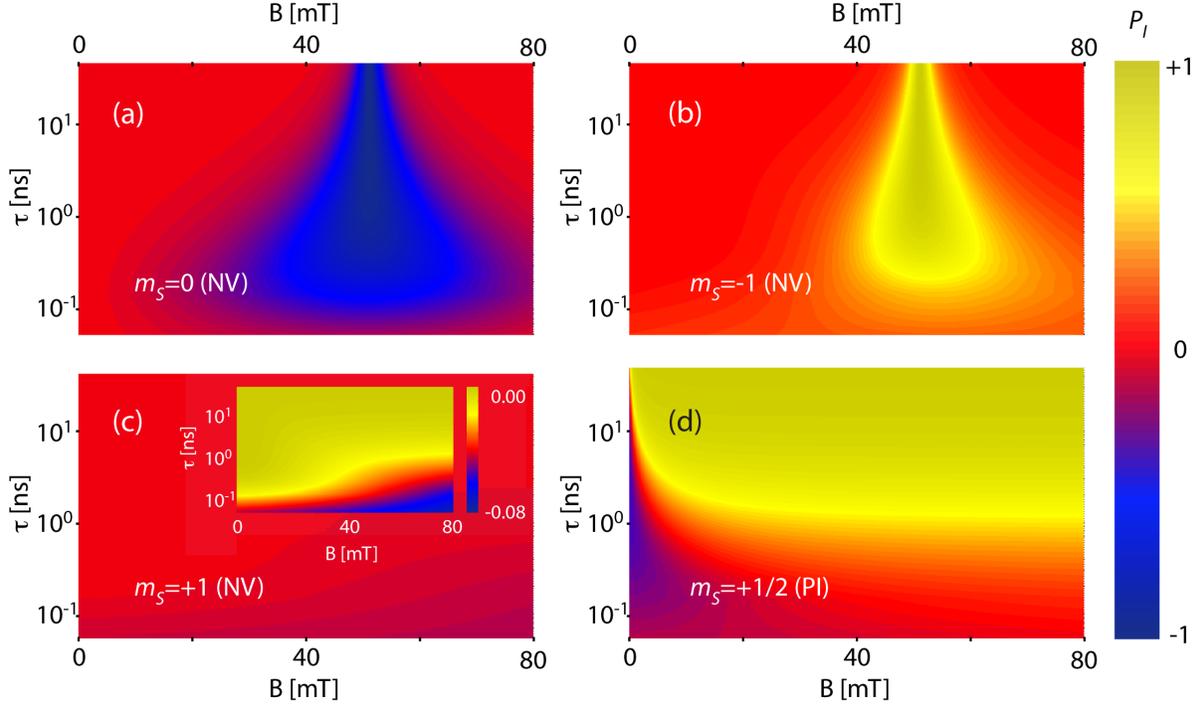

**Fig. 2**: (a) Calculated steady-state spin polarization $P_I \equiv 2\langle I \rangle$ of charge carriers with *g*-factor $g_I = -2$ as a function of the applied magnetic field $B$ and correlation time $\tau_c$ assuming continuous NV pumping into $m_S = 0$. (b,c) Same as in (a) but for the case where the NV is pumped into $m_S = -1$ or $m_S = +1$, respectively. The inset in (c) reproduces the curve in the main graph but using a renormalized color scale. (d) Same as in (a) but for the case where the carrier polarization is induced by P1 centers externally pumped into $m_S = +1/2$. In (a) through (d) the carrier spin-lattice relaxation time is $T_{1I} = 10$ μs, the paramagnetic surface density is $\sigma_{NV} = \sigma_{P1} = 10^{12}$ cm$^{-2}$, and the distance to the conductive layer is $d = 1$ nm.

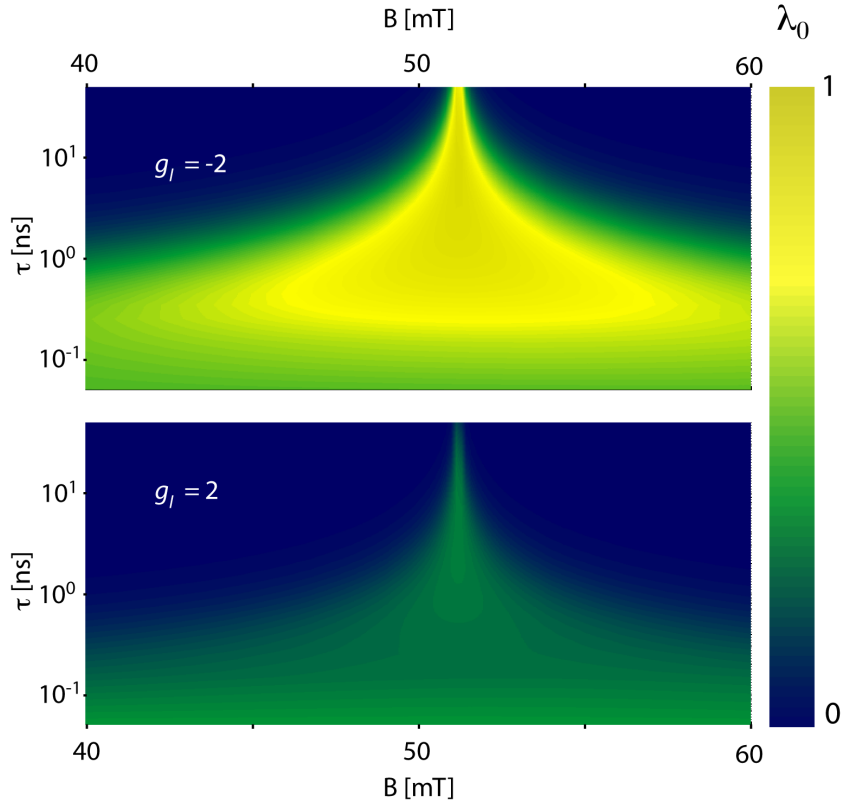

**Fig. 3**: (a) Calculated leakage factor $\lambda_0$ for carriers of g-factor $g_I = -2$ as a function of the applied magnetic field and correlation time. In this case the P1 concentration ($\sigma_{PI} = 10^{12}$ cm$^{-2}$) and distance to the conductive plane ($d = 1$ nm) is the same as for the NV centers; the carrier spin-lattice relaxation time is $T_{1I} = 10$ μs and NVs are pumped into $m_S = 0$. (b) Same as in (a) but for the case where the carrier g-factor is $g_I = 2$.

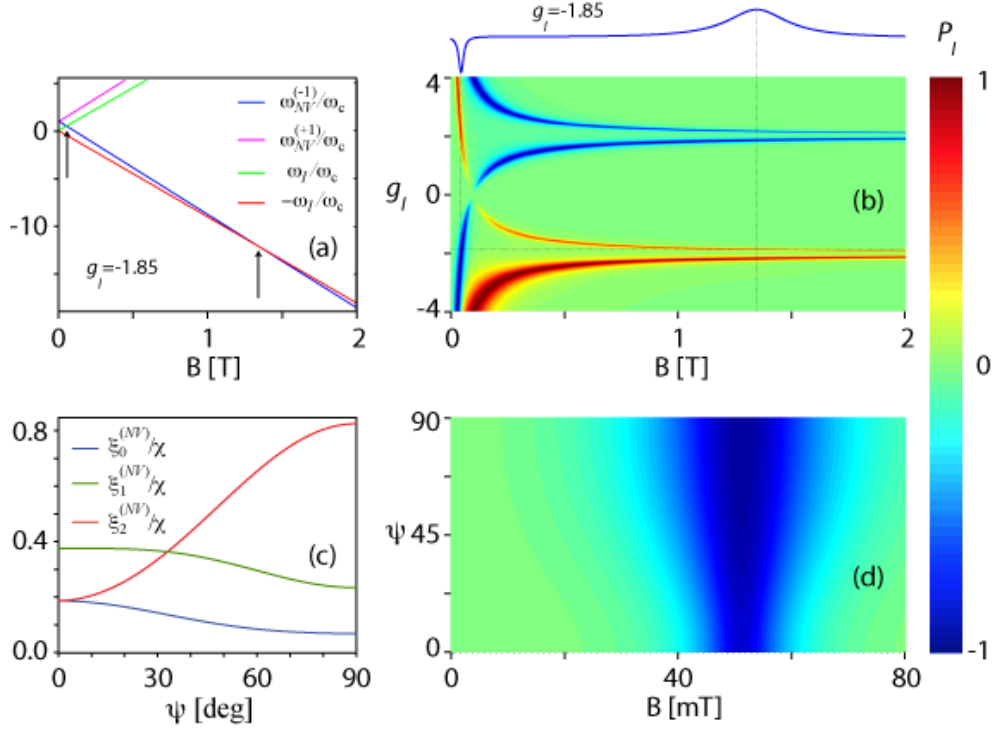

**Fig. 4**: (a) NV- and carrier-spin resonance frequencies (normalized relative to the NV zero-field frequency $\omega_c = 1.87$ GHz) as a function of the magnetic field assuming a carrier g-factor $g_I = 1.85 < 2$. Crossings between $\omega_{NV}^{(-1)}$ and $\omega_I$ or $-\omega_I$ take place at ~0.05 T and ~1.35 T respectively (arrows). (b) NV-induced carrier spin polarization as a function of the carrier g-factor $g_I$ and applied magnetic field $B$ assuming NV pumping into ($m_S = 0$). The upper curve shows the cross section of the graph at $g_I = -1.85$ (horizontal dotted line). The carrier correlation time is $\tau_c = 1$ ns; all other conditions as in Fig. 2. (c) Relative transition probability amplitudes $\xi_j^{(NV)}/\chi$, $j = 0,1,2$ as a function of the angle $\psi$ between the NV symmetry axis and the carrier plane normal. (d) Carrier polarization as a function of the magnetic field $B$ and angle $\theta$ between the NV symmetry axis and the carrier plane normal. The carrier g-factor is $g_I = -2$; all other conditions as in (a).